\documentclass[
  aps,
  prd,
  reprint,
  onecolumn,
  superscriptaddress,
  12pt,
  tightenlines
]{revtex4-2}

\usepackage{amsmath,amssymb,mathtools}
\usepackage{bm}
\usepackage{graphicx}
\usepackage{tikz}
\usetikzlibrary{
  arrows.meta,
  decorations.pathmorphing,
  positioning,
  calc,
  shapes.misc
}
\usepackage{booktabs}

\newcommand{\alphaz}{\alpha Z}

\newcommand{\dd}{\mathrm{d}}
\newcommand{\ii}{\mathrm{i}}
\newcommand{\e}{\mathrm{e}}
\newcommand{\order}{\mathcal{O}}

\makeatletter
\renewcommand\normalsize{%
  \@setfontsize\normalsize{12pt}{14pt}%
}
\makeatother
\begin{document}

\title{Bound-state decay of a charged scalar particle in an external
Coulomb field}

\author{Andrzej Czarnecki}
\author{Artem O. Davydov}

\affiliation{
Department of Physics, University of Alberta,
Edmonton, Alberta T6G 2E1, Canada
}

\begin{abstract}

  We determine the decay rate $\Gamma$ of a scalar particle bound to a
  light nucleus, numerically and analytically. As a model system, we
  consider a negatively charged kaon in the ground state of a
  hydrogen-like ion decaying into charged and neutral scalar pions,
  $K^-\to \pi^-\pi^0$. All particles are assumed pointlike; the
  nucleus is infinitely heavy.  The partial wave expansion converges
  slowly, as in a recent analysis of the bound-muon decay. To overcome
  this difficulty, we introduce a subtraction method: the free $\pi^-$
  contribution is evaluated separately, while the partial-wave sum is
  used only for the Coulomb distortion.  The ratio of bound-to-free
  decay rates, $\Gamma/\Gamma_0$, is found for light nuclei,
  $1\leq Z\leq 10$.  The numerical calculation is complemented by a
  diagrammatic small-$\alphaz$ expansion. We find
  $ \Gamma/\Gamma_0 = 1-(\alpha Z)^2/2 + 17(\alpha Z)^4/24
  +\mathcal{O}((\alpha Z)^5)$, in agreement with the all-orders
  numerical calculation at low $Z$.
\end{abstract}

\maketitle

\section{Introduction}

Decay processes of particles bound in the Coulomb field of a nucleus
probe the interplay of decay dynamics, binding effects, and
final-state interactions.  An example is the decay of a muon
bound in a muonic atom, where the total decay rate differs from that
of a free muon~\cite{Uberall:1960zz,Andreev:2012fj}.  Numerically,
such problems are often formulated through partial-wave expansions of
continuum wave functions~\cite{Watanabe:1993emp,Czarnecki:2026cap,
  Kaygorodov:2025yag,Czarnecki:2025phw,Czarnecki:2011mx}.  For light
nuclei the binding parameter $\alphaz$ is small
($\alpha = 1/137.036$, $Z$ is the atomic number); nevertheless, direct
partial-wave calculations are numerically demanding.  The
angular-momentum series converges slowly. It is costly to compute because each partial wave
involves highly oscillatory radial integrals.

 In a recent calculation of the
Huff factor, the ratio of the bound-muon decay rate to the free-muon
decay rate, Uesaka \emph{et al.} discussed the difficulty of obtaining
stable results for light nuclei. They did not include the region
$Z<6$, where the partial-wave convergence becomes particularly
poor~\cite{Uesaka:2026tdd}.  In the present work we examine this
low-$Z$ convergence problem using a simplified scalar model.
Instead of treating the full spin-$1/2$ problem of bound-muon decay,
we study a scalar analogue in which a negatively charged scalar
particle, denoted by $K^-$, is bound to a static point-like nucleus
and decays into charged and neutral scalar particles,
\begin{equation}
(ZK^-)_{1S}\to Z+\pi^-+\pi^0 .
\end{equation}
The decay is induced by a local three-scalar interaction.  The initial
charged scalar and the outgoing charged pion are treated as point-like
particles described by the Klein--Gordon equation in the nuclear
Coulomb field, with the outgoing pion occupying a continuum scattering
state.

The term ``kaon'' is used as a  label for the initial 
particle.  In an actual kaonic atom, strong-interaction absorption
dominates over weak kaon decay; such effects are not
included in the present model.  Nevertheless, kaonic atoms provide a
context for Coulomb-bound scalar systems. They are of
experimental interest, as illustrated by the recent SIDDHARTA-2
measurement of  strong-interaction effects in kaonic
hydrogen~\cite{Bazzi:2026zmc}.

Here, we use this scalar model to develop methods for bound-state
decays.  The model has common features with the bound-muon decay: an
initial Coulomb bound state and a charged  particle in the final
state. The absence
of spin simplifies the subtraction procedure and the perturbative expansion.  The  small-$\alphaz$ expansion is derived 
diagrammatically through  $(\alphaz)^4$, verifying the low-$Z$ numerical calculation.

The main numerical idea is to subtract the free $\pi^-$
contribution at the level of the radial matrix element.  The free-pion
problem is reduced analytically to a one-dimensional representation,
evaluated separately; the partial-wave sum is applied only to the
Coulomb-distortion correction.  Consequently, the partial-wave
calculation does not have to evaluate the dominant free-pion rate, but
only the much smaller change produced by the final-state Coulomb
interaction.  For $1\le Z\le5$ the two approaches give consistent
results for the Coulomb-distortion correction, while the subtraction
reduces the required partial-wave cutoff by approximately
$25$--$30\%$.  This separation noticeably improves the numerical
efficiency for light nuclei.

The paper is organized as follows.  Section~\ref{sec:model} introduces
the simplified scalar model, with the Klein--Gordon wave functions used
for the bound and continuum states.  Section~\ref{sec:pw} derives the
partial-wave formula for the total decay rate.  Section~\ref{sec:freepion}
derives a one-dimensional
representation of the   free-pion contribution.  In Sec.~\ref{sec:subtraction}  the
subtraction method is  used to isolate the Coulomb-distortion correction to
the final pion state.  Section~\ref{sec:numerics} presents the numerical
results; Sec.~\ref{sec:analytics} derives the diagrammatic
small-$\alphaz$ expansion through order $(\alphaz)^4$.  Finally,
Sec.~\ref{sec:conclusions} summarizes our results.

\section{Scalar-QED model}\label{sec:model}

The nucleus is treated as pointlike, infinitely heavy, and spinless;
QCD effects are neglected. Both final pions are taken to be massless.  The decay
interaction is 
\begin{equation}
    \mathcal{L}_{\rm int}
    =
    g K^-\pi^+\pi^0+\mathrm{h.c.}
\end{equation}
For a charged scalar in an external electrostatic potential energy
$V(r)$, we use the Klein--Gordon equation ($\hbar = c = 1$),
\begin{equation}\label{eq:kg}
    \left[\nabla^2+\left(E-V(r)\right)^2-m^2\right]\phi(\bm r)=0 .
\end{equation}
The potential energy of a particle with charge $-e$ is
\begin{equation}
    V(r)=-\frac{\alphaz}{r}.
\end{equation}
The kaon ground-state wave function is written as
\begin{equation}
    \Psi_K(\bm r)
    =
    N_K^{\rm phys}R_K(r)Y_{00}(\hat r).
\end{equation}
With $M_K,E_K$ denoting the mass and the energy of $K^-$, the Klein--Gordon density normalization is given
by~\cite{Greiner:1990tz} 
\begin{equation}
    \rho_K(\bm r)
    =
    \frac{E_K-V(r)}{M_K}
    |\Psi_K(\bm r)|^2,
    \qquad
    \int \dd^3r\,\rho_K(\bm r)=1 .
    \label{eq:boundnorm}
\end{equation}
Equivalently,
\begin{equation}
    \int_0^\infty \dd r\,r^2
    \left[E_K-V(r)\right]
    \left|N_K^{\rm phys}R_K(r)\right|^2
    =
    M_K .
\end{equation}
For a point Coulomb field, the $1S$  parameters are
\begin{equation}\label{eq:EKgamma}
    E_K=M_K\sqrt{\frac12+\gamma},
    \qquad
    \gamma=\sqrt{\frac14-(\alphaz)^2},
    \qquad
    \beta=2\sqrt{M_K^2-E_K^2}.
\end{equation}
The radial dependence is
\begin{equation}
    R_K(r)=(\beta r)^{\gamma-1/2}\e^{-\beta r/2}.
\end{equation}
In the radial matrix elements below we absorb the angular factor from
the initial $S$-state into the normalization constant,
\begin{equation}\label{eq:NKcode}
    \mathcal{N}_K=\sqrt{4\pi}\,N_K^{\rm phys}.
\end{equation}
Using the exact Coulomb relation
\begin{equation}
    (\alphaz)\beta+E_K(2\gamma+1)=2E_K,
\end{equation}
the point-Coulomb normalization constant can be written in the compact
form
\begin{equation}\label{eq:NK}
    \mathcal{N}_K
    =
    \left[
    \frac{2\pi M_K\beta^3}
    {E_K\,\Gamma(2\gamma+1)}
    \right]^{1/2}.
\end{equation}
The outgoing charged pion is described by an energy-normalized
Klein--Gordon scattering state in the same external Coulomb field.  Its
partial-wave expansion is written as~\cite{Jansen:1987jc}:
\begin{equation}\label{eq:pionpw}
    \phi_{\pi^-}^{(-)*}(\bm r)=
    \sum_{l=0}^{\infty}\sum_{m=-l}^{l}
    Y_{lm}^*(\hat p)Y_{lm}(\hat r)u_{El}(r),
\end{equation}
where $E=p$ for a massless final pion.  The radial functions satisfy
the energy-normalization condition
\begin{equation}\label{eq:contnorm}
    \int_0^\infty \dd r\,r^2
    \left[\frac{E+E'}{2}-V(r)\right]
    u_{El}^*(r)u_{E'l}(r)
    =
    \delta(E-E') .
\end{equation}
For a point Coulomb potential~\cite{Jansen:1987jc}:
\begin{equation}\label{eq:coulrad}
    u_{El}(r)=
    \frac{N_{El}}{r}
    (2\ii pr)^{\mu+1/2}
    \e^{-\ii pr}\,
    {}_1F_1\left(\frac12+\mu+\ii\eta,2\mu+1,2\ii pr\right),
\end{equation}
with
\begin{equation}
    \mu=\sqrt{\left(l+\frac12\right)^2-(\alpha Z)^2},
    \qquad
    \eta=\alpha Z\frac{E}{p}=\alpha Z,
\end{equation}
and
\begin{equation}
    N_{El}
    =
    \frac{\e^{\pi\eta/2}}{\sqrt{2\pi p}}\,
    \frac{\left|\Gamma\left(\frac12+\mu+\ii\eta\right)\right|}
    {\Gamma(2\mu+1)} .
\end{equation}
The $Z\to 0$ limit of Eq.~\eqref{eq:coulrad} defines the free $\pi^-$ wave function in the same hypergeometric phase convention.  It is used in the subtraction method below.

\section{Partial-wave formula}\label{sec:pw}

We now derive the partial-wave expression used for the numerical
calculation.  With the continuum states normalized as in
Eq.~\eqref{eq:contnorm}, the decay width is written as~\cite{Jansen:1987jc}:
\begin{equation}\label{eq:widthmaster}
    \Gamma=
    \frac{2\pi}{M_K}
    \int\frac{\dd^3k}{(2\pi)^3k_0}
    \int \dd E\,\dd\Omega_p\,
    |\mathcal{M}|^2
    \delta(E_K-E-k_0).
\end{equation}
Here $k_0=k$ is the $\pi^0$ energy, and $E=p$ is the $\pi^-$
energy.  The coordinate-space matrix element is
\begin{equation}\label{eq:Mcoord}
    \mathcal{M}
    =
    g\int\dd^3r\,
    \e^{\ii\bm k\cdot\bm r}
    \Psi_K(\bm r)
    \phi_{\pi^-}^{(-)*}(\bm r).
\end{equation}
Expanding the $\pi^0$ plane wave and using the partial-wave
representation of the $\pi^-$ state, the angular integrations
reduce the matrix element to
\begin{equation}\label{eq:Mpw}
    \mathcal{M}
    =
    g\sum_{lm}\ii^l
    Y_{lm}^*(\hat k)Y_{lm}(\hat p)S_l(k,E),
\end{equation}
where the radial amplitude is
\begin{equation}\label{eq:Sl}
    S_l(k,E)
    =
    \mathcal{N}_K
    \int_0^\infty \dd r\,r^2
    j_l(kr)R_K(r)u_{El}(r).
\end{equation}
The angular orthogonality gives
\begin{equation}
    \int \dd\Omega_k\dd\Omega_p\,|\mathcal{M}|^2
    =
    g^2\sum_{l=0}^\infty(2l+1)|S_l(k,E)|^2 .
\end{equation}
Using the energy delta function with $k=k_0=E_K-E$, we obtain
\begin{equation}\label{eq:ratioPW}
    \frac{\Gamma}{\Gamma_0}
    =
    \frac{1}{2\pi}
    \sum_{l=0}^\infty(2l+1)
    \int_0^{E_K}\dd k\,k\,
    |S_l(k,E_K-k)|^2 .
\end{equation}
The free kaon decay rate is
\begin{equation}\label{eq:Gamma0}
    \Gamma_0=\frac{g^2}{2\pi M_K}.
\end{equation}
Equation~\eqref{eq:ratioPW} is the basic all-orders  formula in
our scalar-QED model.

\section{Free-pion contribution}\label{sec:freepion}

Here we derive the bound-$K$ decay rate, describing the outgoing $\pi^-$ by a plane-wave (neglecting final-state interactions); our subtraction method use this result.

We denote the  free
$\pi^-$ radial function by $u^{(0)}_{El}$, the
$Z\to0$ limit of  $u_{El}$
in Eq.~\eqref{eq:coulrad}. 
 The free partial-wave
amplitude is
\begin{equation}\label{eq:S0def}
    S_l^{(0)}(k,E)
    =
    \mathcal{N}_K
    \int_0^\infty \dd r\,r^2\,
    j_l(kr)R_K(r)u^{(0)}_{El}(r).
\end{equation}
Substituting this free limit into Eq.~\eqref{eq:ratioPW} gives
\begin{equation}\label{eq:freePW}
    \left(\frac{\Gamma}{\Gamma_0}\right)_{\mathrm{free}\,\pi}=
    \frac{1}{2\pi}\sum_{l=0}^\infty(2l+1)
    \int_0^{E_K}\dd k\,k\,|S_l^{(0)}(k,E_K-k)|^2 .
\end{equation}
The free-pion contribution can also be written avoiding partial waves.  The matrix element is proportional to the
Fourier transform of the bound-state wave function,
\begin{equation}\label{eq:freeAmp}
    {\cal M}_0(\bm p,\bm k)
    =
    g\int \dd^3r\,
    e^{i(\bm p+\bm k)\cdot\bm r}\,
    \Psi_K(\bm r)
    \equiv
    g\widetilde\Psi_K(\bm q),
    \qquad
    \bm q=\bm p+\bm k .
\end{equation}
After the two-body phase-space integration, the free-pion decay-rate
ratio becomes
\begin{equation}\label{eq:freeMomentumGeneral}
    \left(\frac{\Gamma}{\Gamma_0}\right)_{\mathrm{free}\,\pi}
    =
    \frac{4\pi}{(2\pi)^3}
    \int_0^{E_K}\dd q\,q^2\,
    \left|\widetilde\Psi_K(q)\right|^2 .
\end{equation}
For the point-Coulomb $1S$ Klein--Gordon wave function this reduces to
the single integral
\begin{align}
    \left(\frac{\Gamma}{\Gamma_0}\right)_{\mathrm{free}\,\pi}
    &=
    4\pi\frac{\mathcal{N}_K^2}{(2\pi)^3}
    \int_0^{E_K}\dd q\,
    \left[
    \frac{
    2^{\gamma+3/2}\Gamma(\gamma+3/2)
    }{
    \beta^2
    }
    \left(1+\frac{4q^2}{\beta^2}\right)^{-(2\gamma+3)/4}
    \right.
    \nonumber\\
    &\hspace{3.6cm}\left.
    \times
    \sin\left[
    \frac{2\gamma+3}{2}
    \arctan\left(\frac{2q}{\beta}\right)
    \right]
    \right]^2 .
    \label{eq:freeMomentum1D}
\end{align}
This  representation gives a separate numerical
evaluation of the free-pion contribution to all orders in binding, see Table~\ref{tab:finalnum} below.

\section{Subtraction method}\label{sec:subtraction}

Equation~\eqref{eq:ratioPW} becomes inefficient at small $Z$.  In this
region the interacting-pion decay rate is very close to the
corresponding free-pion result, while the Coulomb-distortion
correction is much smaller than either of them.  A partial-wave
calculation must therefore reconstruct the full decay rate with
sufficient absolute accuracy to resolve this small correction.  The
same numerical difficulty appears in realistic bound-muon
calculations.  Uesaka \emph{et al.}~recently calculated the Huff
factor for isotopes with $6\le Z\le 94$~\cite{Uesaka:2026tdd}.  They
omitted lighter nuclei, $Z<6$, because mean-field calculations are
less reliable for light nuclei and because the partial-wave
convergence of the Huff factor becomes poor in this region.  The
present scalar model allows us to access this low-$Z$ regime.

The interacting radial function is written as
\begin{equation}
    u_{El}(r)=u^{(0)}_{El}(r)+\Delta u_{El}(r),
\end{equation}
which induces the corresponding decomposition of the partial-wave
amplitude,
\begin{equation}
    S_l(k,E)=S_l^{(0)}(k,E)+\Delta S_l(k,E).
\end{equation}
Here $S_l^{(0)}$ is the free-pion amplitude of
Eq.~\eqref{eq:S0def};
\begin{equation}\label{eq:DeltaSdef}
    \Delta S_l(k,E)
    =
    \mathcal{N}_K
    \int_0^\infty \dd r\,r^2\,
    j_l(kr)R_K(r)
    \left[u_{El}(r)-u^{(0)}_{El}(r)\right].
\end{equation}
The subtraction is performed point by point under the radial integral.
Thus $\Delta S_l$ is a genuine Coulomb-distortion amplitude.

The full decay rate is written as
\begin{equation}\label{eq:subrate}
    \frac{\Gamma}{\Gamma_0}
    =
    \left(\frac{\Gamma}{\Gamma_0}\right)_{\mathrm{free}\,\pi}
    +\Delta_C,
\end{equation}
where $\Delta_C$ is the difference between the interacting-pion and
free-pion decay-rate ratios.
Using $S_l=S_l^{(0)}+\Delta S_l$,  it can be computed as
\begin{equation}\label{eq:deltac}
    \Delta_C
    =
    \frac{1}{2\pi}\sum_{l=0}^{\infty}(2l+1)
    \int_0^{E_K}\dd k\,k\,
    \left[
    |\Delta S_l|^2
    +
    2\operatorname{Re}\left(\Delta S_l S_l^{(0)*}\right)
    \right].
\end{equation}
The numerical advantage of Eq.~\eqref{eq:deltac} comes from evaluating
the dominant free-pion contribution separately from the one-dimensional
representation in Eq.~\eqref{eq:freeMomentum1D}, while the partial-wave
sum is applied  to the much smaller Coulomb-distortion
correction $\Delta_C$.  Thus the calculation avoids reconstructing a
decay-rate ratio close to unity with the high  accuracy required
to resolve the small Coulomb correction.

\section{Numerical results}\label{sec:numerics}

We evaluate the bound-kaon decay-rate ratio $\Gamma/\Gamma_0$ using
the separation introduced in Eq.~\eqref{eq:subrate}.  The free-pion
contribution is computed separately from the one-dimensional expression
in Eq.~\eqref{eq:freeMomentum1D}, to all orders in $\alphaz$.  The
Coulomb-distortion correction $\Delta_C$ is computed from the
subtracted partial-wave sum in Eq.~\eqref{eq:deltac}, using the
interacting $\pi^-$ Klein--Gordon wave function.  The results for
kaons bound to light point-like nuclei with charges $1\le Z\le10$ are
shown in Table~\ref{tab:finalnum}.  We present the rate suppression,
$1-\Gamma/\Gamma_0$, rather than the rate itself.

\begin{table}[htb]
\centering
\caption{Numerical results for the decay-rate ratio $\Gamma/\Gamma_0$
of a kaon bound to a light point-like nucleus.  Column 1 gives
the nuclear charge $Z$.  Column 2 shows the free-pion rate
suppression, $1-(\Gamma/\Gamma_0)_{\mathrm{free}\,\pi}$ from
Eq.~\eqref{eq:freeMomentum1D}.  Column 3 gives the
Coulomb-distortion correction to the decay rate,
$\Delta_C=\Gamma/\Gamma_0-(\Gamma/\Gamma_0)_{\mathrm{free}\,\pi}$, see  Eq.~\eqref{eq:deltac}.  Column 4
shows the interacting-pion rate suppression, $1-\Gamma/\Gamma_0$,
including the Coulomb distortion of $\pi^-$.  Since
$\Delta_C$ is a positive correction to the rate, column 4 is
obtained from columns 2 and 3 as
$1-\Gamma/\Gamma_0=1-(\Gamma/\Gamma_0)_{\mathrm{free}\,\pi}-\Delta_C$.  Column 5
shows $\Delta_C$ in units of $(\alpha Z)^4$; the last column gives
the partial-wave cutoff in the subtracted calculation.}
\label{tab:finalnum}
\begin{tabular}{c@{\hspace{2em}} c@{\hspace{2em}} c@{\hspace{2em}} c@{\hspace{2em}} c@{\hspace{2em}} c}
\toprule
$Z$ &
$10^{5}\,[1-(\Gamma/\Gamma_0)_{\mathrm{free}\,\pi}]$ &
$10^{9}\,\Delta_C$ &
$10^{5}\,(1-\Gamma/\Gamma_0)$ &
$\Delta_C/(\alpha Z)^4$ &
$l_{\max}$ \\
\midrule
1  & 2.6627 & 3.817 & 2.6624 & 1.346 & 802 \\
2  & 10.653 & 61.70 & 10.647 & 1.360 & 384 \\
3  & 23.979 & 315.8 & 23.947 & 1.375 & 238 \\
4  & 42.651 & 1009  & 42.550 & 1.390 & 162 \\
5  & 66.690 & 2494  & 66.440 & 1.407 & 133 \\
6  & 96.119 & 5236  & 95.596 & 1.425 & 113 \\
7  & 130.97 & 9829  & 129.99 & 1.444 & 98 \\
8  & 171.29 & 17000 & 169.59 & 1.463 & 76 \\
9  & 217.13 & 27620 & 214.37 & 1.484 & 69 \\
10 & 268.55 & 42720 & 264.28 & 1.507 & 62 \\
\bottomrule
\end{tabular}
\end{table}

The reduced $\pi^-$ radial function
$F_{El}(r)\equiv r u_{El}(r)$ is obtained by numerically solving
the radial Klein--Gordon equation.  Near the origin, the regular
solution is written as
\begin{equation}
F_{El}(r)=A r^s g(r), \qquad
s=\mu+\frac12,\qquad
\mu=\sqrt{\left(l+\frac12\right)^2-(\alpha Z)^2},
\end{equation}
so that the known regular power-law behavior is factored out.  The function $g(r)$ is initialized from its regular Frobenius
expansion. It is integrated with the adaptive eighth-order Runge--Kutta
method DOP853~\cite{Hairer:1993}.  The interacting and $Z=0$ radial
functions are integrated as components of the same
ODE system; they are evaluated on the same adaptively
generated radial mesh.  This is useful for the point-by-point
subtraction in Eq.~\eqref{eq:DeltaSdef}, where the difference between
two close radial solutions is required.  Beyond the centrifugal
scale, the integration is continued  for $F_{El}(r)$;
the radial function entering the matrix element is recovered as
$u_{El}(r)=F_{El}(r)/r$.

The radial and momentum integrals in
Eqs.~(\ref{eq:DeltaSdef},\ref{eq:deltac}) are done using
composite Gauss--Legendre quadrature.  Their numerical stability was
checked by varying the number of integration intervals and quadrature
points, as well as the ODE tolerances.

The main numerical challenge is the convergence of the partial-wave
expansion.  To estimate the uncertainty associated with truncating the
series, the last terms of the partial-wave sum were approximated by a
geometric progression; the remaining tail was summed analytically,
as in Ref.~\cite{Czarnecki:2025phw}.  For each  $Z$,
we chose $l_{\max}$ such that the estimated tail was smaller than
one half of the last quoted digit of $\Delta_C$.  The resulting
values of $l_{\max}$ are listed in Table~\ref{tab:finalnum}.  The
stability of the results was also checked by extending the partial-wave
sums beyond these cutoffs and by varying the numerical integration
parameters.

To quantify the numerical gain from the subtraction, we also performed
partial-wave calculations of the full interacting-pion decay
rate for $1\le Z\le5$.  This is the low-$Z$ region not
included in Ref.~\cite{Uesaka:2026tdd}.  Our scalar model
allows us to  examine this numerical difficulty without
additional uncertainty from nuclear modeling.  The geometric-tail
criterion was applied to both the direct and subtracted calculations.  The
corresponding partial-wave cutoffs are shown in
Table~\ref{tab:convergence}.

\begin{table}[htb]
\centering
\caption{Partial-wave cutoffs in the direct and subtracted
calculations for the five lowest-$Z$ nuclei.  
The  large-$l$ tails were estimated by approximating omitted
terms by a geometric progression, with the cutoff chosen to stabilize
the four quoted significant digits of $\Delta_C$.}
\label{tab:convergence}
\begin{tabular}{c@{\hspace{2em}} c@{\hspace{2em}} c}
\toprule
$Z$ & $l_{\max}^{\rm direct}$ & $l_{\max}^{\rm subtracted}$ \\
\midrule
1 & 1136 & 802 \\
2 & 526  & 384 \\
3 & 323  & 238 \\
4 & 221  & 162 \\
5 & 177  & 133 \\
\bottomrule
\end{tabular}
\end{table}

For all five nuclei, the values of $\Delta_C$ obtained from the direct
and subtracted calculations agree in all four quoted significant
digits.  The subtraction reduces the required partial-wave cutoff by
approximately $25$--$30\%$.  More importantly, the subtracted
calculation evaluates the small correction $\Delta_C$, whereas the
direct approach obtains it from the difference between the free- and
interacting-pion decay-rate ratios.  Since both ratios are close to
unity, the direct approach requires their more precise evaluation to
reach the same accuracy for $\Delta_C$.

\begin{figure*}[htb]
\centering
\includegraphics[width=0.48\textwidth]
{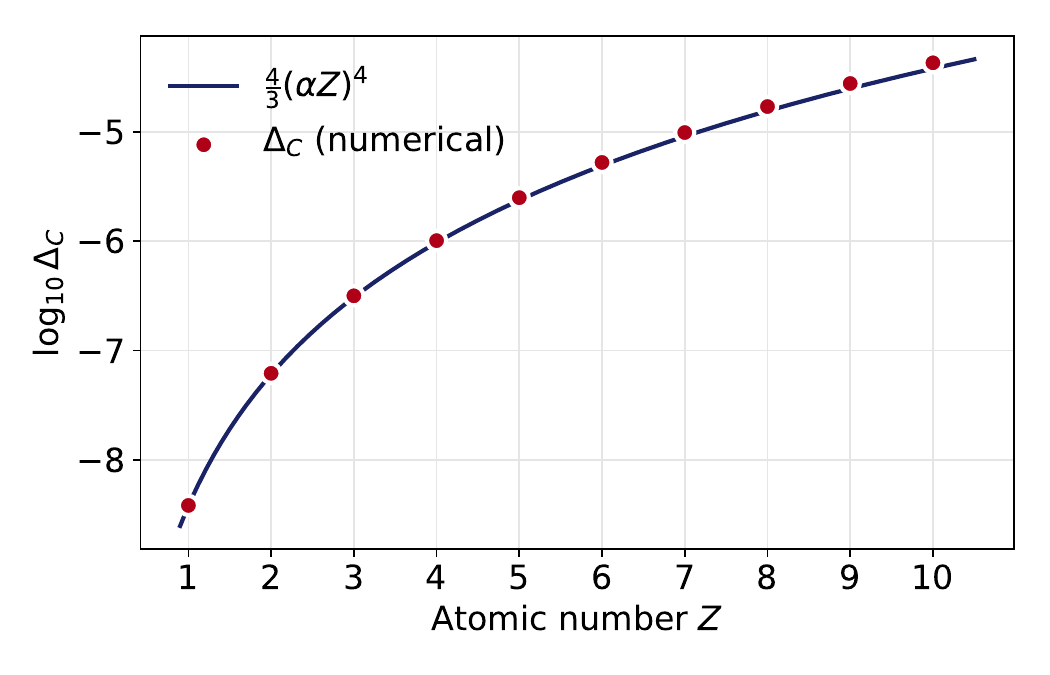}
\hfill
\includegraphics[width=0.48\textwidth]
{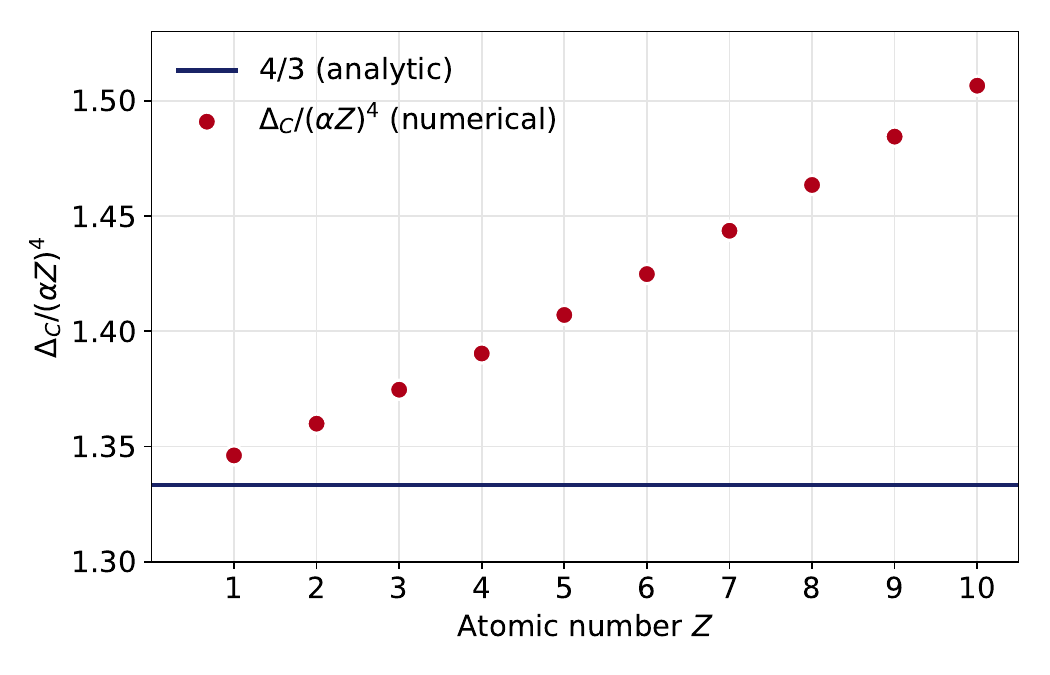}
\caption{Coulomb-distortion correction in the bound kaon decay.   
$\Delta_C$ is the difference between the normalized decay rate $\Gamma/\Gamma_0$
obtained with the interacting and free outgoing $\pi^-$.
Left panel: log of the absolute correction
$\Delta_C$. Right panel:  the normalized correction
$\Delta_C/(\alpha Z)^4$.  The dots are the numerical
results.  In the left graph, the solid curve is the log of the leading analytic
prediction $\Delta_C=(4/3)(\alpha Z)^4$; in the right graph, the
horizontal line marks the coefficient $4/3$.
As $Z$ decreases, the normalized numerical results approach $4/3$,
confirming the leading $(\alpha Z)^4$ behavior.  The deviations at
larger $Z$ arise from higher-order contributions.}
\label{fig:deficitZ}
\end{figure*}

Figure~\ref{fig:deficitZ} compares the all-orders numerical
Coulomb-distortion correction with the leading analytic result.  The
left graph shows the absolute correction $\Delta_C$, while the right
graph shows $\Delta_C/(\alpha Z)^4$.  As $Z$ decreases, the
normalized numerical results approach the analytic coefficient $4/3$,
in agreement with the leading-order result derived in
Sec.~\ref{sec:analytics}.  The deviations at larger $Z$ are due to
higher-order contributions beyond the leading $(\alpha Z)^4$ term.

As derived in Sec.~\ref{sec:analytics}, the Coulomb distortion of the
outgoing charged pion does not modify the leading free-pion
suppression, which starts at order $(\alpha Z)^2$, but first
contributes to the decay rate at order $(\alpha Z)^4$.  The total
suppression therefore has the leading form
\begin{equation}
1-\frac{\Gamma}{\Gamma_0} =
\frac12(\alpha Z)^2+\cdots .
\end{equation}
This coefficient coincides with the leading \"Uberall result for
bound-muon decay, but it arises differently.  In the
nonrelativistic point-nucleus treatment of bound-muon decay, the
plane-wave-electron contribution is $1-(9/2)(\alpha Z)^2+\cdots$,
while the Coulomb distortion of the outgoing electron contributes
$+4(\alpha Z)^2$, producing the net coefficient
$-1/2$~\cite{Uberall:1960zz,Uesaka:2026tdd}.  In the present scalar
model the free-pion contribution alone gives
$1-(\alpha Z)^2/2+\cdots$, and the Coulomb distortion of the
outgoing pion starts at order $(\alpha Z)^4$.

The $(\alpha Z)^4$ term in Eq.~\eqref{eq:finalexpansion} is 
the first nontrivial correction to the leading $(\alpha Z)^2$
suppression in the scalar problem.  Its inclusion accounts for the
dominant Coulomb distortion seen in the numerical results. The
remaining deviation at larger $Z$ is due to higher-order
terms.

\section{Small-$\alpha Z$ expansion}\label{sec:analytics}

The low-$Z$ numerical calculation can be checked analytically by expanding the decay rate in powers of $\alphaz$.  Coulomb interactions with the nucleus are represented by insertions of the external potential
\begin{equation}
    V_C(\bm Q)=-\frac{4\pi \alpha Z}{Q^2},
    \qquad Q=|\bm Q| .
\end{equation}
This is the Fourier transform of the coordinate-space potential
$V(r)=-\alpha Z/r$ used in the Klein--Gordon equation. The charged scalar couples through the usual scalar-QED one-photon and seagull vertices.

We formulate the calculation in terms of the imaginary part
of the kaon self-energy in the external Coulomb field. With the external
kaon described by the momentum-space bound-state wave function
$\widetilde\Psi_K$, we write
\begin{equation}\label{eq:selfenergyMaster}
    \frac{\Gamma}{\Gamma_0}
    =
    \int\frac{\dd^3q}{(2\pi)^3}
    \frac{\dd^3q'}{(2\pi)^3}
    \widetilde\Psi_K^*(\bm q')\,
    {\cal K}(\bm q',\bm q)\,
    \widetilde\Psi_K(\bm q).
\end{equation}
The $\pi^-\pi^0$ loop kernel is
\begin{equation}\label{eq:Kdef}
    {\cal K}(\bm q',\bm q)
    \equiv
    -\frac{8}{\Gamma_0}\,
    \mathrm{Im}\,\Sigma(E_K;\bm q',\bm q) .
\end{equation}
Here $\Sigma(E_K;\bm q',\bm q)$ denotes the kaon self-energy
pion loop in the external field. The factor of 8 results from the bound- and continuum-state
normalization 
\cite{Greiner:1990tz,Jansen:1987jc}.

We compute the kernel perturbatively in the external Coulomb field,
\begin{equation}
    {\cal K}
    =
    {\cal K}_0
    +
    {\cal K}_{1\gamma}
    +
    {\cal K}_{2\gamma}
    +\cdots .
\end{equation}
The terms ${\cal K}_0$, ${\cal K}_{1\gamma}$, and
${\cal K}_{2\gamma}$ are generated by the pion-loop self-energy with
zero, one, and two Coulomb photon insertions on the $\pi^-$ line,
respectively. For the coefficients
through order $(\alpha Z)^4$, the finite end point $E_K$ of the physical
momentum integral can be replaced by infinity without changing the
result at this accuracy.

In this section and Appendix~\ref{app:onephoton}, the factors
of $i$ arising from the Feynman rules are combined into the overall
prefactors.  The  $i$ relating the Minkowski
loop integrals to the Passarino--Veltman functions is displayed below.

The order of each contribution follows from bound-state power
counting.  The loop momentum is hard, $l\sim M_K$, while the
Coulomb momenta are of the order of the bound-state scale
$\kappa=M_K\alpha Z$.  Expanding the loop in the Coulomb momenta
produces local operators constructed from the potential and its
gradients, with
\begin{equation}
    V\sim\alpha Z\kappa\sim M_K(\alpha Z)^2,
    \qquad
    \nabla\sim\kappa .
\end{equation}
A term with $n$ Coulomb-field insertions, without gradients, is
nominally of order $(\alpha Z)^{2n}$, while each pair of gradients
introduces an additional factor of $(\alpha Z)^2$.  Complete
gauge-invariant contributions without gradients vanish: a constant
shift of the potential only shifts the kaon energy, on which
$\operatorname{Im}\Sigma$ does not depend for massless pions, apart
from the end-point correction of order $(\alpha Z)^5$.  The
one-insertion contribution therefore starts with the operator
$\nabla^2V$, at order $(\alpha Z)^4$.  At two insertions, the terms
proportional to $V^2$, which are also nominally of order
$(\alpha Z)^4$, cancel in the complete gauge-invariant result, as
verified  below.  Contributions with three or more
Coulomb-field insertions start at order $(\alpha Z)^6$; they 
do not affect the decay rate through order $(\alpha Z)^4$.

This power counting is consistent with the results obtained
below:
\begin{equation}
    \frac{\langle V^2\rangle}{M_K^2}
    =
    2(\alpha Z)^4,
    \qquad
    \frac{\langle\nabla^2V\rangle}{3M_K^3}
    =
    \frac{4}{3}(\alpha Z)^4 .
\end{equation}

In each case we first write the loop integral contributing to
$\Sigma(E_K;\bm q',\bm q)$.  The tensor integrals are reduced to
scalar master integrals by integration-by-parts (IBP)
identities~\cite{Chetyrkin:1981qh,Laporta:2000dsw}, using
Kira~3~\cite{Lange:2025fba}. 
We then take the imaginary part of the reduced result and expand the corresponding kernel, organizing the resulting terms by the order in $\alpha Z$ at which they contribute to the decay rate.
 Finally, the expanded kernel is inserted into
Eq.~\eqref{eq:selfenergyMaster}, and the remaining momentum integrals
are evaluated.

\subsection{Zero Coulomb photons}

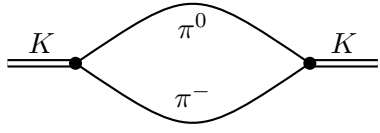
\begin{figure}[tbp]
\centering
\begin{tikzpicture}[line width=0.9pt, scale=1.0]
\tikzset{scalar/.style={thick},kaon/.style={double,double distance=1.6pt,thick}}
\coordinate (L) at (-1.55,0);
\coordinate (R) at (1.55,0);
\draw[kaon] (-2.45,0) -- (L) node[midway,above] {$K$};
\draw[kaon] (R) -- (2.45,0) node[midway,above] {$K$};
\filldraw (L) circle (2pt);
\filldraw (R) circle (2pt);
\draw[scalar] (L) .. controls (0,1.05) .. (R);
\draw[scalar] (L) .. controls (0,-1.05) .. (R);
\node[fill=white,inner sep=1.0pt] at (0,0.50) {$\pi^0$};
\node[fill=white,inner sep=1.0pt] at (0,-0.43) {$\pi^-$};
\end{tikzpicture}
\caption{Zeroth-order kaon self-energy diagram.  The double line
denotes the external kaon, the single lines denote $\pi^-$ and $\pi^0$. The imaginary part is
obtained from the two-pion cut.}
\label{fig:selfenergy0}
\end{figure}

The zeroth-order contribution is the kaon self-energy diagram shown in
Fig.~\ref{fig:selfenergy0}.  Its imaginary part is generated by the
two-pion cut.

Since there are no external Coulomb-field insertions, no
three-momentum is transferred to the kaon by the decay kernel.
The zero-photon kernel is therefore diagonal in the kaon momentum,
\begin{equation}
    {\cal K}_0(\bm q',\bm q)
    =
    (2\pi)^3\delta^3(\bm q'-\bm q)\,
    {\cal K}_0^{\rm red}(\bm q).
\end{equation}
We define its reduced diagonal part by
\begin{equation}\label{eq:K0loop}
    {\cal K}_0^{\rm red}(\bm q)
    =
    -\frac{8}{\Gamma_0}\,
    \operatorname{Im} I_0^{\rm red}(\bm q).
\end{equation}
With the normalization of the bound-state kernel fixed by
Eq.~\eqref{eq:boundnorm}, the reduced zeroth-order self-energy
contribution is
\begin{equation}\label{eq:I0loop}
    I_0^{\rm red}(\bm q)
    =
    -\frac{g^2}{16\pi^2M_K}\,B_0(s),
    \qquad
    s\equiv E_K^2-\bm q^{\,2}.
\end{equation}
For the two-massless-pion cut,
\begin{equation}\label{eq:ImB0}
    \operatorname{Im}B_0(s)
    =
    \pi\,\theta(s).
\end{equation}
Therefore
\begin{equation}\label{eq:ImI0}
    \operatorname{Im}I_0^{\rm red}(\bm q)
    =
    -\frac{\Gamma_0}{8}\,
    \theta(E_K^2-\bm q^{\,2}),
\end{equation}
where
\begin{equation}
    \Gamma_0=\frac{g^2}{2\pi M_K}.
\end{equation}
The full zero-photon kernel is thus
\begin{equation}\label{eq:K0identity}
    {\cal K}_0(\bm q',\bm q)
    =
    (2\pi)^3\delta^3(\bm q'-\bm q)\,
    \theta(E_K^2-\bm q^{\,2}).
\end{equation}
Substitution into Eq.~\eqref{eq:selfenergyMaster} gives
\begin{equation}\label{eq:zerophotonEKM}
    \left(\frac{\Gamma}{\Gamma_0}\right)_{\mathrm{free}\,\pi}
    =
    \int\frac{\dd^3q}{(2\pi)^3}
    |\widetilde\Psi_K(\bm q)|^2
    \theta(E_K^2-\bm q^{\,2}).
\end{equation}
If the momentum integral is extended to infinity, the
Klein--Gordon normalization gives
\begin{equation}
    \int\frac{\dd^3q}{(2\pi)^3}
    |\widetilde\Psi_K(\bm q)|^2
    =
    \int\dd^3r\,|\Psi_K(\bm r)|^2
    =
    \frac{E_K}{M_K}.
\end{equation}
The difference between this expression and
Eq.~\eqref{eq:zerophotonEKM} is the high-momentum tail
$|\bm q|>E_K$.  Using the leading Coulomb $1S$ wave function, this
tail starts at order $(\alpha Z)^5$,
\begin{equation}
   \left(\frac{\Gamma}{\Gamma_0}\right)_{\mathrm{free}\,\pi}
    =
    \frac{E_K}{M_K}
    -
    \frac{32}{5\pi}(\alpha Z)^5
    +\cdots .
\end{equation}
Hence, through order $(\alpha Z)^4$,
\begin{equation}\label{eq:freeexpDiag}
    \left(\frac{\Gamma}{\Gamma_0}\right)_{\mathrm{free}\,\pi}
    =
    1-\frac12(\alphaz)^2-\frac58(\alphaz)^4
    +\order\left((\alphaz)^5\right).
\end{equation}

\subsection{One Coulomb photon}

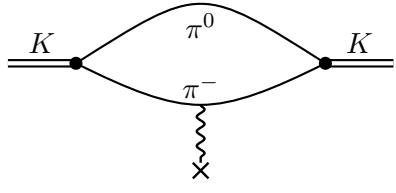
\begin{figure}[tbp]
\centering
\begin{tikzpicture}[line width=0.9pt, scale=1.0]
\tikzset{photon/.style={decorate,decoration={snake,amplitude=1.45pt,segment length=5.5pt,pre length=0pt,post length=0pt},line cap=butt},scalar/.style={thick},kaon/.style={double,double distance=1.6pt,thick},insertion/.style={cross out,draw,minimum size=5pt,inner sep=0pt}}
\coordinate (L) at (-1.65,0);
\coordinate (R) at (1.65,0);
\draw[kaon] (-2.55,0) -- (L) node[midway,above] {$K$};
\draw[kaon] (R) -- (2.55,0) node[midway,above] {$K$};
\filldraw (L) circle (2pt);
\filldraw (R) circle (2pt);
\coordinate (P) at (0.00,-0.55);
\draw[scalar] (L) .. controls (-0.95,-0.35) and (-0.45,-0.55) .. (P)
              .. controls (0.45,-0.55) and (0.95,-0.35) .. (R);
\draw[scalar] (L) .. controls (0,1.05) .. (R);
\node[fill=white,inner sep=1.0pt] at (0,0.50) {$\pi^0$};
\node[fill=white,inner sep=1.0pt] at (0.00,-0.31) {$\pi^-$};
\node[insertion] (X) at (0.00,-1.42) {};
\draw[photon] (P) -- (X);
\end{tikzpicture}
\caption{One-Coulomb-photon insertion on the $\pi^-$ line. The cross denotes the external Coulomb field.}
\label{fig:selfenergy1}
\end{figure}

The one-photon correction, shown in
Fig.~\ref{fig:selfenergy1}, is the first perturbative insertion of the
external Coulomb field on the $\pi^-$ line.

We define
\begin{equation}\label{eq:K1def}
    {\cal K}_{1\gamma}(\bm q_1)
    =
    -\frac{8}{\Gamma_0}\,\mathrm{Im}\,I_{1\gamma}(\bm q_1).
\end{equation}
The one-photon contribution is
\begin{equation}\label{eq:I1main}
    I_{1\gamma}(\bm q_1)
    =
    \frac{i g^2}{M_K}\,V_C(\bm q_1)
    \int\frac{\dd^4 l}{(2\pi)^4}
    \frac{2l_0}{D_0D_1D_2}
    =
    -\frac{g^2}{16\pi^2M_K}\,
    V_C(\bm q_1)\,
    {\cal C}_{1\gamma}.
\end{equation}
Here
\begin{equation}
    D_0=l^2+\ii0,
    \qquad
    D_1=(q-l)^2+\ii0,
    \qquad
    D_2=(l-q_1)^2+\ii0 ,
\end{equation}
where $q^\mu$ is the four-momentum carried by the external kaon line
and $q_1^\mu=(0,\bm q_1)$ is the Coulomb momentum transfer.  The
Coulomb potential is
\begin{equation}
    V_C(\bm q_1)=-\frac{4\pi\alpha Z}{\bm q_1^{\,2}} .
\end{equation}
The reduced numerator integral ${\cal C}_{1\gamma}$ is evaluated in
Appendix~\ref{app:onephoton}.  The leading term in the kernel expansion
is
\begin{equation}\label{eq:K1result}
    {\cal K}_{1\gamma}(\bm q_1)
    =
    \frac{4\pi \alpha Z}{3M_K^3} + {\cal{O}}((\alphaz)^3).
\end{equation}
This kernel is independent of $\bm q_1=\bm q'-\bm q$. The corresponding
particular case of Eq.~\eqref{eq:selfenergyMaster} is
\begin{align}
    \left(\frac{\delta\Gamma}{\Gamma_0}\right)_{1\gamma}
    &=
    \int\frac{\dd^3q}{(2\pi)^3}
    \frac{\dd^3q'}{(2\pi)^3}
    \widetilde\Psi_K^*(\bm q')
    {\cal K}_{1\gamma}(\bm q'-\bm q)
    \widetilde\Psi_K(\bm q) \notag\\
    &=
    \frac{4\pi \alpha Z}{3M_K^3}
    \int\frac{\dd^3q}{(2\pi)^3}
    \frac{\dd^3q'}{(2\pi)^3}
    \widetilde\Psi_K^*(\bm q')
    \widetilde\Psi_K(\bm q)
    =
    \frac{4\pi \alpha Z}{3M_K^3}\,|\Psi_K(0)|^2 .
\end{align}
To obtain the coefficient of $(\alpha Z)^4$, it is sufficient to
replace the bound-state wave function by its leading Schrödinger
$1S$ form,
\begin{equation}
    \Psi_K(\bm r)\to \Psi_{1S}(\bm r),
    \qquad
    |\Psi_{1S}(0)|^2=\frac{\kappa^3}{\pi}.
\end{equation}
Thus
\begin{equation}\label{eq:onephoton}
    \left(\frac{\delta\Gamma}{\Gamma_0}\right)_{1\gamma}
    =
    \frac{4}{3}(\alphaz)^4 .
\end{equation}

\subsection{Two Coulomb photons}

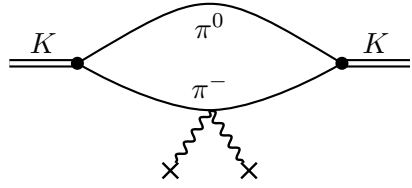
\begin{figure}[tbp]
\centering
\begin{tikzpicture}[line width=0.9pt, scale=1.0]
\tikzset{
photon/.style={decorate,decoration={snake,amplitude=1.45pt,
segment length=5.5pt,pre length=0pt,post length=0pt},line cap=butt},
scalar/.style={thick},
kaon/.style={double,double distance=1.6pt,thick},
insertion/.style={cross out,draw,minimum size=5pt,inner sep=0pt}}
\coordinate (L) at (-1.75,0);
\coordinate (R) at (1.75,0);
\draw[kaon] (-2.65,0) -- (L) node[midway,above] {$K$};
\draw[kaon] (R) -- (2.65,0) node[midway,above] {$K$};
\filldraw (L) circle (2pt);
\filldraw (R) circle (2pt);
\coordinate (SG) at (0,-0.62);
\draw[scalar] (L) .. controls (-1.15,-0.35) and (-0.45,-0.62) .. (SG)
              .. controls (0.45,-0.62) and (1.15,-0.35) .. (R);
\draw[scalar] (L) .. controls (0,1.05) .. (R);
\node[fill=white,inner sep=1.0pt] at (0,0.50) {$\pi^0$};
\node[fill=white,inner sep=1.0pt] at (0,-0.30) {$\pi^-$};
\node[insertion] (X1) at (-0.52,-1.42) {};
\node[insertion] (X2) at (0.52,-1.42) {};
\draw[photon] (SG) -- (X1);
\draw[photon] (SG) -- (X2);
\end{tikzpicture}
\caption{Scalar-QED seagull insertion.  Both Coulomb photons attach to
the same $\pi^-$ vertex.}
\label{fig:selfenergy2sg}
\end{figure}

\begin{figure}[tbp]
\centering
\begin{tikzpicture}[line width=0.9pt, scale=1.0]
\tikzset{
photon/.style={decorate,decoration={snake,amplitude=1.45pt,
segment length=5.5pt,pre length=0pt,post length=0pt},line cap=butt},
scalar/.style={thick},
kaon/.style={double,double distance=1.6pt,thick},
insertion/.style={cross out,draw,minimum size=5pt,inner sep=0pt}}
\coordinate (L) at (-1.75,0);
\coordinate (R) at (1.75,0);
\draw[kaon] (-2.65,0) -- (L) node[midway,above] {$K$};
\draw[kaon] (R) -- (2.65,0) node[midway,above] {$K$};
\filldraw (L) circle (2pt);
\filldraw (R) circle (2pt);
\coordinate (P1) at (-0.70,-0.55);
\coordinate (P2) at (0.70,-0.55);
\draw[scalar] (L) .. controls (-1.15,-0.35) and (-0.95,-0.55) .. (P1)
              .. controls (-0.25,-0.78) and (0.25,-0.78) .. (P2)
              .. controls (0.95,-0.55) and (1.15,-0.35) .. (R);
\draw[scalar] (L) .. controls (0,1.05) .. (R);
\node[fill=white,inner sep=1.0pt] at (0,0.50) {$\pi^0$};
\node[fill=white,inner sep=1.0pt] at (0,-0.38) {$\pi^-$};
\node[insertion] (X1) at (-0.70,-1.42) {};
\node[insertion] (X2) at (0.70,-1.42) {};
\draw[photon] (P1) -- (X1);
\draw[photon] (P2) -- (X2);
\end{tikzpicture}
\caption{Two Coulomb-field insertions on the $\pi^-$ line.
The crosses denote the external Coulomb field.}
\label{fig:selfenergy2lad}
\end{figure}
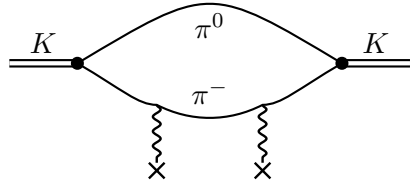

At second order in the external Coulomb field there are two
contributions: the scalar-QED seagull insertion shown in
Fig.~\ref{fig:selfenergy2sg} and the contribution with two linear
insertions of the external field on the $\pi^-$ line shown in
Fig.~\ref{fig:selfenergy2lad}.  At this order we need only the leading
nonrelativistic $1S$ wave function.

\subsubsection{Seagull graph}

The seagull graph contains the scalar-QED two-photon vertex on the
$\pi^-$ line.  We define its contribution to the kernel by
\begin{equation}\label{eq:KsgDef}
    {\cal K}_{2\gamma}^{\rm sg}(\bm q',\bm q)
    =
    -\frac{8}{\Gamma_0}\,
    \operatorname{Im} I_{\rm sg}(\bm q',\bm q),
    \qquad
    Q=|\bm q'-\bm q| .
\end{equation}
The loop integral has the form
\begin{equation}\label{eq:IsgMain}
    I_{\rm sg}(\bm q',\bm q)
    =
    g^2\frac{2\pi^2(\alpha Z)^2}{Q}\,
    \frac{1}{16\pi^2M_K}\,
    C_0,
    \qquad
    Q=|\bm q'-\bm q| .
\end{equation}
Here the factor $2\pi^2(\alpha Z)^2/Q$ comes from the convolution of
the two Coulomb potentials, while $C_0$ denotes the  scalar
triangle master in the Passarino--Veltman
normalization~\cite{Passarino:1978jh}.  The leading term of this master integral required for the kernel expansion is
\begin{equation}\label{eq:ImC0SeagullLO}
    \operatorname{Im} C_0
    =
    -\frac{\pi}{M_K^2}
    +{\cal O}((\alpha Z)^2).
\end{equation}
Using this result in Eq.~\eqref{eq:KsgDef}, and the normalization of
Eq.~\eqref{eq:Gamma0}, we obtain
\begin{equation}\label{eq:Ksgresult}
    {\cal K}_{2\gamma}^{\rm sg}(\bm q',\bm q)
    =
    \frac{g^2(\alpha Z)^2\pi}
    {\Gamma_0 M_K^3|\bm q'-\bm q|}.
\end{equation}
At the required order we use the leading nonrelativistic Coulomb
$1S$ wave function in momentum space,
\begin{equation}
    \widetilde\Psi_K(\bm q)
    =
    N\,\widetilde\Psi_{1S}(\bm q),
    \qquad
    \widetilde\Psi_{1S}(\bm q)
    =
    \frac{8\pi\kappa}{(\bm q^2+\kappa^2)^2},
\end{equation}
where
\begin{equation}
    N^2=\frac{\kappa^3}{\pi}.
\end{equation}
Substitution into Eq.~\eqref{eq:selfenergyMaster} gives
\begin{align}
    \left(\frac{\delta\Gamma_2}{\Gamma_0}\right)_{\rm sg}
    &=
    \frac{g^2(\alpha Z)^2N^2\pi}{\Gamma_0 M_K^3}
    \int\frac{\dd^3q}{(2\pi)^3}
    \frac{\dd^3q'}{(2\pi)^3}
    \widetilde\Psi_{1S}^{\,*}(\bm q')
    \frac{1}{|\bm q'-\bm q|}
    \widetilde\Psi_{1S}(\bm q).
\end{align}
Introducing the momentum transfer
\begin{equation}
    \bm Q=\bm q'-\bm q,
\end{equation}
the double integral becomes
\begin{equation}
    \int\frac{\dd^3Q}{(2\pi)^3}\frac{1}{Q}
    \int\frac{\dd^3q}{(2\pi)^3}
    \widetilde\Psi_{1S}^{\,*}(\bm q+\bm Q)
    \widetilde\Psi_{1S}(\bm q).
\end{equation}
The inner convolution is
\begin{equation}
    \int\frac{\dd^3q}{(2\pi)^3}
    \widetilde\Psi_{1S}^{\,*}(\bm q+\bm Q)
    \widetilde\Psi_{1S}(\bm q)
    =
    \frac{16\pi\kappa}
    {(Q^2+4\kappa^2)^2}.
\end{equation}
Hence
\begin{align}
    \int\frac{\dd^3Q}{(2\pi)^3}\frac{1}{Q}
    \int\frac{\dd^3q}{(2\pi)^3}
    \widetilde\Psi_{1S}^{\,*}(\bm q+\bm Q)
    \widetilde\Psi_{1S}(\bm q)
    &=
    \frac{8\kappa}{\pi}
    \int_0^\infty
    \frac{Q\,\dd Q}{(Q^2+4\kappa^2)^2}
    = \frac{1}{\pi\kappa}.
\end{align}
Therefore, using $N^2=\kappa^3/\pi$ and
$\Gamma_0=g^2/(2\pi M_K)$, we obtain
\begin{equation}\label{eq:sgWidthIntegral}
    \left(\frac{\delta\Gamma_2}{\Gamma_0}\right)_{\rm sg}
    =
    2(\alpha Z)^4 .
\end{equation}

\subsubsection{Two Coulomb insertions}

The second two-Coulomb-photon contribution arises from two linear
insertions of the external Coulomb field on the $\pi^-$ line.
We define its contribution to the kernel by
\begin{equation}\label{eq:KladDef}
    {\cal K}_{2\gamma}^{\rm lad}(\bm q',\bm q)
    =
    -\frac{8}{\Gamma_0}\,
    \operatorname{Im} I_{\rm lad}(\bm q',\bm q),
    \qquad
    Q=|\bm q'-\bm q| .
\end{equation}
The two-photon contribution has the form
\begin{equation}\label{eq:KladLoop}
    I_{\rm lad}(\bm q',\bm q)
    =
    g^2\int\frac{\dd^3k}{(2\pi)^3}
    V_C(\bm k)V_C(\bm Q-\bm k)\,
    \frac{1}{16\pi^2M_K}\,
    I_{\rm lad}^{\rm loop}.
\end{equation}
The four-momenta of the two Coulomb-field insertions are
\begin{equation}
    q_1^\mu=(0,\bm k),
    \qquad
    q_2^\mu=(0,\bm Q-\bm k),
    \qquad
    (q_1+q_2)^\mu=(0,\bm Q).
\end{equation}
The corresponding loop integral is defined by
\begin{equation}\label{eq:ladderint}
    I_{\rm lad}^{\rm loop}
    =
    16\pi^2 i
    \int\frac{\dd^4 l}{(2\pi)^4}
    \frac{4l_0^2}{D_0D_1D_2D_3},
\end{equation}
with
\begin{equation}
    D_3=(l-q_1-q_2)^2+\ii0 .
\end{equation}
The numerator is reduced by introducing
\begin{equation}
    D_4=(l+n)^2,
    \qquad
    n=(M_K,\bm0),
    \qquad
    l_0=\frac{D_4-D_0-M_K^2}{2M_K}.
\end{equation}
After reducing the resulting scalar integrals by IBP identities, the
leading term of the loop integral required for the kernel expansion is
\begin{equation}\label{eq:IladResult}
    \operatorname{Im} I_{\rm lad}^{\rm loop}
    =
    \frac{\pi}{M_K^2}
    +{\cal O}\!\left((\alpha Z)^2\right).
\end{equation}
The Coulomb-field convolution gives
\begin{equation}\label{eq:VVconvolution}
    \int\frac{\dd^3k}{(2\pi)^3}
    V_C(\bm k)V_C(\bm Q-\bm k)
    =
    (4\pi\alpha Z)^2
    \int\frac{\dd^3k}{(2\pi)^3}
    \frac{1}{\bm k^{\,2}(\bm Q-\bm k)^2}
    =
    \frac{2\pi^2(\alpha Z)^2}{Q}.
\end{equation}
Substituting Eqs.~\eqref{eq:IladResult} and
\eqref{eq:VVconvolution} into Eqs.~\eqref{eq:KladLoop} and
\eqref{eq:KladDef}, we obtain
\begin{equation}\label{eq:Kladresult}
    {\cal K}_{2\gamma}^{\rm lad}(\bm q',\bm q)
    =
    -\frac{g^2(\alpha Z)^2\pi}
    {\Gamma_0 M_K^3|\bm q'-\bm q|}.
\end{equation}
This kernel is equal in magnitude and opposite in sign to the seagull
kernel in Eq.~\eqref{eq:Ksgresult}.  Therefore the same bound-state
convolution evaluated above gives
\begin{align}
    \left(\frac{\delta\Gamma_2}{\Gamma_0}\right)_{\rm lad}
    &=
    -\frac{g^2(\alpha Z)^2N^2\pi}{\Gamma_0 M_K^3}
    \int\frac{\dd^3Q}{(2\pi)^3}\frac{1}{Q}
    \int\frac{\dd^3q}{(2\pi)^3}
    \widetilde\Psi_{1S}^{\,*}(\bm q+\bm Q)
    \widetilde\Psi_{1S}(\bm q)
    \notag\\
    &=
    -2(\alpha Z)^4 .
\end{align}
The seagull and two-insertion contributions  cancel in the
decay rate at order $(\alpha Z)^4$,
\begin{equation}
    \left(\frac{\delta\Gamma_2}{\Gamma_0}\right)_{\rm sg}
    +
    \left(\frac{\delta\Gamma_2}{\Gamma_0}\right)_{\rm lad}
    =
    0 .
\end{equation}
The final-state Coulomb correction at this order is therefore the
one-photon result of Eq.~\eqref{eq:onephoton}.  Combining
Eq.~\eqref{eq:onephoton} with the free-pion result
Eq.~\eqref{eq:freeexpDiag}, we obtain
\begin{equation}\label{eq:finalexpansion}
    \frac{\Gamma}{\Gamma_0}
    =
    1-\frac12(\alpha Z)^2
    +\frac{17}{24}(\alpha Z)^4
    +{\cal O}\!\left((\alpha Z)^5\right).
\end{equation}
The leading coefficient of the Coulomb-distortion correction,
\begin{equation}
    \Delta_C
    =
    \frac{4}{3}(\alpha Z)^4
    +{\cal O}\!\left((\alpha Z)^5\right),
\end{equation}
obtained in this section, is confirmed by the numerical results in
Table~\ref{tab:finalnum}, where
$\Delta_C/(\alpha Z)^4$ approaches $4/3$ at small $Z$.

\section{Conclusion}\label{sec:conclusions}

We have studied the decay of a scalar bound
to a low-$Z$ point-like nucleus as a simplified model of bound-state decay
in an external Coulomb field.  The model was formulated in scalar QED:
the nucleus was treated as infinitely heavy and spinless, the initial
charged scalar, referred to as a kaon, and the final charged pion were
described by Klein--Gordon wave functions in the Coulomb field, and both
final particles were taken to be massless.  Recoil, nuclear
polarization, finite-size, and QCD effects were neglected.  In this
approximation only the external Coulomb field corrects the
decay rate.

We derived a partial-wave formula for the total decay rate,
valid to all orders in $\alphaz$.  Its  numerical evaluation,
however, becomes inefficient for low-$Z$ nuclei.  At small $Z$, the
decay rates obtained with the interacting and free $\pi^-$ wave
functions are very close, so a partial-wave calculation must
have
 sufficient absolute accuracy to
resolve the  small Coulomb-distortion correction.

To avoid this difficulty, we introduced a subtraction method.  The
free-pion problem is reduced analytically to a one-dimensional
representation.  The partial-wave sum is then applied to the
Coulomb-distortion correction $\Delta_C$, obtained from the difference
between the interacting $\pi^-$ wave function and its $Z\to0$ limit.
This reorganization avoids reconstructing the dominant free
contribution through the partial-wave expansion.  Direct and
subtracted calculations for $1\le Z\le5$ agree in all four quoted
significant digits. The subtraction reduces the number of partial
waves by $25$--$30\%$.  More importantly, the subtracted calculation
evaluates the small Coulomb-distortion correction directly rather than
as the difference between two decay-rate ratios close to unity.  The
resulting numerical improvement is especially useful at $Z=1$, where
the Coulomb correction to the free-pion result is particularly small.

We also derived the small-$\alphaz$ expansion through order
$(\alphaz)^4$ using Feynman diagrams in a static nuclear field.  The
free-pion contribution gives the coefficient $-5/8$ at order
$(\alphaz)^4$, while the Coulomb distortion of the outgoing charged
pion does not modify the leading $(\alphaz)^2$ term and first
contributes at order $(\alphaz)^4$.  The one-photon contribution gives
$+(4/3)(\alphaz)^4$.  The two-photon seagull and
 two-insertion contributions, $+2(\alphaz)^4$ and
$-2(\alphaz)^4$, sum to zero.  Combining the
free-pion and Coulomb-distortion contributions, we obtain
\begin{equation}
\frac{\Gamma}{\Gamma_0}
=
1-\frac12(\alphaz)^2
+\frac{17}{24}(\alphaz)^4
+\order\left((\alphaz)^5\right).
\end{equation}

The all-orders numerical calculation agrees with this analytic
expansion for light nuclei, with
$\Delta_C/(\alpha Z)^4$ approaching $4/3$ as $Z$ decreases.
The  numerical comparison for $1\le Z\le5$ also demonstrates
the effectiveness of the subtraction method in the low-$Z$ region
where partial-wave convergence is particularly demanding.  Although
the leading coefficient $-1/2$ coincides with the \"Uberall result
for bound-muon decay, its origin is different in the present scalar
model: the Coulomb distortion of the outgoing pion starts only at
order $(\alphaz)^4$.  The subtraction strategy and the diagrammatic
approach developed here provide  tools for future
calculations of bound-state decays in external Coulomb fields,
including the more realistic spin-$1/2$ problem of bound-muon decay.

\begin{acknowledgments}
This work was supported by the Natural Sciences and Engineering
Research Council of Canada (NSERC).
\end{acknowledgments}

\appendix

\section{One-photon master-integral reduction}\label{app:onephoton}

This appendix gives the details of the reduction leading to
Eq.~\eqref{eq:K1result}.  We start from the reduced numerator integral
${\cal C}_{1\gamma}$ defined in Eq.~\eqref{eq:I1main}.  Introducing the standard Passarino--Veltman tensor triangle
coefficient $C^{\mu}$, we write
\begin{equation}
    \int\frac{\dd^4 l}{(2\pi)^4}
    \frac{l^\mu}{D_0D_1D_2}
    =
    \frac{i}{16\pi^2}\,C^{\mu}
    =
    \frac{i}{16\pi^2}
    \left(q^\mu C_q+q_1^\mu C_1\right).
\end{equation}
The denominators are defined in Sec.~\ref{sec:analytics}.
Since $q_1^0=0$, the numerator $2l_0$ projects only onto the
coefficient $C_q$.  To the accuracy needed for the
$(\alpha Z)^4$ coefficient,
$q^0=E_K=M_K+\order((\alpha Z)^2)$, and therefore
\begin{equation}
    {\cal C}_{1\gamma}=2M_K C_q .
\end{equation}
Let
\begin{equation}
    s=q^2,\qquad
    u=q_1^2,\qquad
    P=q\cdot q_1,\qquad
    t=(q-q_1)^2=s+u-2P .
\end{equation}
Inverting the Gram matrix gives
\begin{equation}
    C_q=
    \frac{u(q\cdot C)-P(q_1\cdot C)}
    {su-P^2}.
\end{equation}
The contractions reduce the tensor integral to scalar bubble and
triangle master integrals.  We use the shorthand notation
\begin{equation}
B_0(x)\equiv B_0(x;0,0),
\qquad
C_0\equiv C_0(s,t,u;0,0,0).
\end{equation}
Here all internal pion masses are zero.  With this convention,
\begin{align}
    q\cdot C
    &=
    \frac12\left[B_0(t)-B_0(u)+sC_0\right],\\
    q_1\cdot C
    &=
    \frac12\left[B_0(t)-B_0(s)+uC_0\right].
\end{align}
Therefore
\begin{equation}\label{eq:CpmastersApp}
    C_q=
    \frac{(u-P)B_0(t)+PB_0(s)-uB_0(u)+u(s-P)C_0}
    {2(su-P^2)} .
\end{equation}
In the physical region relevant for the two-pion cut,
\begin{equation}
    s>0,\qquad t>0,\qquad u<0,
\end{equation}
so that
\begin{equation}
    \operatorname{Im}B_0(s)
    =
    \operatorname{Im}B_0(t)
    =
    \pi,
    \qquad
    \operatorname{Im}B_0(u)=0 .
\end{equation}
Defining
\begin{equation}
    K=\sqrt{P^2-su},
    \qquad
    A=s-P,
\end{equation}
the triangle master gives
\begin{equation}
    \operatorname{Im}C_0
    =
    -\frac{\pi}{2K}
    \ln\frac{A+K}{A-K}.
\end{equation}
Substitution into Eq.~\eqref{eq:CpmastersApp} yields
\begin{equation}\label{eq:ImCpApp}
    \operatorname{Im}C_q
    =
    -\frac{\pi u}{2K^2}
    +\frac{\pi uA}{4K^3}
    \ln\frac{A+K}{A-K}.
\end{equation}
Keeping the leading term required for the kernel expansion and using
$u=-\bm q_1^{\,2}$, we obtain
\begin{equation}
    \operatorname{Im}C_q
    =
    -\frac{\pi\bm q_1^{\,2}}{6M_K^4},
\end{equation}
and hence
\begin{equation}
    \operatorname{Im}{\cal C}_{1\gamma}
    =
    -\frac{\pi\bm q_1^{\,2}}{3M_K^3}.
\end{equation}
Using Eq.~\eqref{eq:I1main}, we reproduce the kernel  in Eq.~\eqref{eq:K1result},
\begin{equation}
    \operatorname{Im}I_{1\gamma}(\bm q_1)
    =
    -\frac{g^2\alpha Z}{12M_K^4}
    =
    -\frac{\pi\Gamma_0}{6M_K^3}\alpha Z .
\end{equation}


\begin{thebibliography}{10}
\providecommand{\url}[1]{\texttt{#1}}
\providecommand{\urlprefix}{URL }
\providecommand{\eprint}[2][]{\url{#2}}

\bibitem{Uberall:1960zz}
H.~\"Uberall, \emph{{Decay of $\mu^-$ Mesons Bound in the K Shell of Light
  Nuclei}}, Phys. Rev. \textbf{119}, 365--376 (1960).

\bibitem{Andreev:2012fj}
V.~Andreev et~al., \emph{{Measurement of Muon Capture on the Proton to 1\%
  Precision and Determination of the Pseudoscalar Coupling $g_P$}},
  Phys.~Rev.~Lett. \textbf{110}, 012504 (2013), \eprint{1210.6545}.

\bibitem{Watanabe:1993emp}
R.~Watanabe, K.~Muto, T.~Oda, T.~Niwa, H.~Ohtsubo, R.~Morita, and M.~Morita,
  \emph{Asymmetry and energy spectrum of electrons in bound-muon decay}, Atomic
  Data and Nucl. Data Tables \textbf{54}, 165 (1993).

\bibitem{Czarnecki:2026cap}
A.~Czarnecki and A.~O. Davydov, \emph{{Radiative Corrections in Bound States:
  Recent Results}}, in \emph{{Loops and Legs in Quantum Field Theories}}
  (2026), \eprint{2606.27600}.

\bibitem{Kaygorodov:2025yag}
M.~Y. Kaygorodov, Y.~S. Kozhedub, A.~V. Malyshev, A.~O. Davydov, Y.~Wu, and
  S.~B. Zhang, \emph{{Study of atomic effects on electron spectrum in
  bound-muon decay process}}, Chin. Phys. C \textbf{50}, 063103 (2026),
  \eprint{2506.02416}.

\bibitem{Czarnecki:2025phw}
A.~Czarnecki, A.~O. Davydov, and M.~Y. Kaygorodov, \emph{{Total decay rate of a
  muon bound to a light nucleus}}, Phys. Rev. D \textbf{113}, 036028 (2026),
  \eprint{2512.23023}.

\bibitem{Czarnecki:2011mx}
A.~Czarnecki, X.~Garcia~i Tormo, and W.~J. Marciano, \emph{{Muon decay in
  orbit: spectrum of high-energy electrons}}, Phys. Rev. \textbf{D84}, 013006
  (2011), \eprint{1106.4756}.

\bibitem{Uesaka:2026tdd}
Y.~Uesaka, T.~Naito, S.~Ebata, and M.~Niikura, \emph{Comprehensive table of
  calculated Huff factors}, Atomic Data and Nuclear Data Tables page 101809
  (2026), arXiv:2602.07501.

\bibitem{Bazzi:2026zmc}
M.~Bazzi et~al., \emph{{High-precision measurement of the kaonic hydrogen 1s
  level shift and width with SIDDHARTA-2}}  (2026), \eprint{2607.13952}.

\bibitem{Greiner:1990tz}
W.~Greiner, \emph{Relativistic Quantum Mechanics. Wave Equations}, Springer,
  Berlin, 3rd edition (2000).

\bibitem{Jansen:1987jc}
G.~Jansen, M.~Pusch, and G.~Soff, \emph{Continuum solutions of the Klein-Gordon
  equation}, Zeitschrift f{\"u}r Physik D Atoms, Molecules and Clusters
  \textbf{8}, 315--327 (1988).

\bibitem{Hairer:1993}
E.~Hairer, S.~P. N{\o}rsett, and G.~Wanner, \emph{Solving ordinary differential
  equations I: Nonstiff problems}, Springer, Berlin (1993).

\bibitem{ParticleDataGroup:2026mpi}
F.~Takahashi et~al., \emph{{Review of Particle Physics}}, Int. J. Mod. Phys. A
  \textbf{41}, 2630011 (2026).

\bibitem{Chetyrkin:1981qh}
K.~G. Chetyrkin and F.~V. Tkachev, \emph{Integration by parts: the algorithm to
  calculate {$\beta$-}functions in 4 loops}, Nucl. Phys. \textbf{B192},
  159--204 (1981).

\bibitem{Laporta:2000dsw}
S.~Laporta, \emph{{High-precision calculation of multiloop Feynman integrals by
  difference equations}}, Int. J. Mod. Phys. A \textbf{15}, 5087--5159 (2000),
  \eprint{hep-ph/0102033}.

\bibitem{Lange:2025fba}
F.~Lange, J.~Usovitsch, and Z.~Wu, \emph{{Kira 3: integral reduction with
  efficient seeding and optimized equation selection}}, Comput. Phys. Commun.
  \textbf{322}, 109999 (2026), \eprint{2505.20197}.

\bibitem{Passarino:1978jh}
G.~Passarino and M.~J.~G. Veltman, \emph{{One Loop Corrections for $e^+ e^-$
  Annihilation Into $\mu^+ \mu^-$ in the Weinberg Model}}, Nucl. Phys. B
  \textbf{160}, 151--207 (1979).

\end{thebibliography}

\end{document}